# Orbital misalignment of the Neptune-mass exoplanet GJ 436b with the spin of its cool star.

Vincent Bourrier[1], Christophe Lovis[1], Hervé Beust[2], David Ehrenreich[1], Gregory W. Henry[3], Nicola Astudillo-Defru[1], Romain Allart[1], Xavier Bonfils[2], Damien Ségransan[1], Xavier Delfosse[2], Heather M. Cegla[1], Aurélien Wyttenbach[1], Kevin Heng[4], Baptiste Lavie[1], Francesco Pepe[1]

**The angle between the spin of a star and its planets' orbital planes traces the history of the planetary system. Exoplanets orbiting close to cool stars are expected to be on circular, aligned orbits because of strong tidal interactions with the stellar convective envelope[1]. Spin-orbit alignment can be measured when the planet transits its star, but such ground-based spectroscopic measurements are challenging for cool, slowly-rotating stars[2]. Here we report the characterization of a planet three-dimensional trajectory around an M dwarf star, derived by mapping the spectrum of the stellar photosphere along the chord transited by the planet[3]. We find that the eccentric orbit of the Neptune-mass exoplanet GJ 436b is nearly perpendicular to the stellar equator. Both eccentricity and misalignment, surprising around a cool star, can result from dynamical interactions (via Kozai migration[4]) with a yet-undetected outer companion. This inward migration of GJ 436b could have triggered the atmospheric escape that now sustains its giant exosphere[5]. Eccentric, misaligned exoplanets orbiting close to cool stars might thus hint at the presence of unseen perturbers and illustrate the diversity of orbital architectures seen in exoplanetary systems.**

Three transits of GJ 436b, which occur every 2.64 days[2], were observed on 9 May 2007 (visit 1)[2], 18 March 2016 (visit 2) and 11 April 2016 (visit 3) with the HARPS (visit 1) and HARPS-N (visits 2–3) spectrographs[6,7]. All visits cover the full transit duration, with exposure times of 300–400 s, and provide baselines of 3–8 h before or after the transit. We corrected spectra for the variability in the distribution of their flux with wavelength caused by Earth's atmosphere (Methods) before using a binary mask to calculate cross-correlation functions (CCFs) that represent an average of the spectral lines from the M dwarf host GJ 436. We introduce a double-Gaussian model to accurately fit the distinctive CCF profiles of M dwarfs (Extended Data Figs 1 and 2) and to improve the stability and precision of their derived contrast, width and radial velocity (RV). These properties show little dispersion around their average values in each visit and are stable between the HARPS-N visits, in agreement with the low activity of GJ 436[2,8] (Extended Data Fig 3).

[1]Observatoire de l'Université de Genève, 51 chemin des Maillettes, 1290 Versoix, Switzerland. [2]Univ. Grenoble Alpes, CNRS, IPAG, F-38000 Grenoble, France. [3]Center of Excellence in Information Systems, Tennessee State University, Nashville, TN 37209, USA. [4]University of Bern, Center for Space and Habitability, Sidlerstrasse 5, CH-3012, Bern, Switzerland.

The observed CCFs originate from starlight integrated over the disk of GJ 436 ($CCF_{DI}$). During the transit they are deprived of the light from the planet-occulted regions ($CCF_{PO}$), which we retrieve using the reloaded Rossiter-McLaughlin (RM) technique[3]. $CCF_{DI}$ are shifted into the star rest frame, then co-added and continuum-normalized outside of the transit to build a master-out template $CCF_{DI}^{OT}$ for each visit. In-transit $CCF_{DI}$ are continuum-scaled according to the depth of the light curve derived from high-precision photometry[2], before subtracting them from the $CCF_{DI}^{OT}$ to retrieve the $CCF_{PO}$ (Methods). The local stellar line profile from the spatially-resolved region of the photosphere occulted by GJ 436b along the transit chord is clearly detected in the $CCF_{PO}$ (Fig. 1, Extended Data Fig. 4). We applied a double-Gaussian model to $CCF_{PO}$ to derive their properties, linking the profiles of the Gaussian components in the same way as for the $CCF_{DI}$ (Methods). We kept in our analysis $CCF_{PO}$ where the stellar line contrast is detected at more than $5\sigma$. Excluded $CCF_{PO}$ (Extended Data Table 1) are faint, associated to darker regions of the stellar limb only partially occulted by GJ 436b. The RV centroids of the $CCF_{PO}$ directly trace the velocity field of the stellar photosphere (Extended Data Fig 5). The three series of surface RVs are consistent over most of the transit (even though they were obtained with two instruments over a 9-year interval) and are predominantly positive (showing that GJ 436b occults redshifted regions of the stellar disk rotating away from us and excluding an aligned system). We simultaneously fitted the three RVs series with the reloaded RM model[3], using a Metropolis-Hasting Markov Chain Monte Carlo (MCMC) algorithm[9] and assuming a solid-body rotation for the star (Methods). The model then depends on the sky-projected obliquity $\lambda_b$ (the angle between the projected angular momentum vectors of the star and of the orbit of GJ436b) and projected rotational velocity $V_{eq} \sin i_\star$ (with $i_\star$ the inclination of the star spin axis relative to our line-of-sight). The best fit (Fig. 1, Extended Data Fig. 5) matches visits 1–2 well, and it yields a relatively large $\chi^2$ of 42 for 19 degrees of freedom because three measurements in Visit 3 deviate by 2.5–3$\sigma$. Excluding them yields $\chi_r^2$=1.1 and does not change the derived properties beyond their 1$\sigma$ uncertainties (Methods), therefore they were kept in the final fit. Posterior probability distributions of the MCMC parameters (Extended Data Fig. 6) are well defined and yield $V_{eq} \sin i_\star = 330^{+90}_{-70}$ m s$^{-1}$ (>190 m s$^{-1}$ with 99% confidence) and $\lambda_b = 72^{+33}_{-24}$° (> 30° with 99% confidence). These properties do not change beyond their 1σ uncertainties when system parameters are varied within their error bars. The Bayesian information criterion (BIC) for the best-fit solid-body model (48) is much lower than for a null velocity model (74) and an aligned model (88). The M dwarf GJ 436 is thus the coolest star across which the RM effect has been detected, with a highly-misaligned orbit for its Neptune-mass companion (Fig. 2).

The slow rotation of GJ 436 is consistent with published upper limits[2,10]. It yields a small amplitude of 1.3 m s$^{-1}$ for the classical RV anomaly - much smaller than the stellar surface velocities measured with the reloaded RM technique - which could not be detected in earlier analysis of Visit 1[2]. The widths of the CCF$_{PO}$ show little dispersion around the width of the CCF$_{DI}^{OT}$, consistent with the non-detection of rotational broadening (Extended Data Fig. 5). The three visits show similar properties for the CCF$_{PO}$ along the transit chord and for the CCF$_{DI}^{OT}$, consistent with the low activity of GJ 436[11,12] and stable emission at ultraviolet[5], optical[8], and infrared[2,13] wavelengths. Nonetheless, small periodic variations in its visible flux[8] and the periodic modulation we measure in HARPS[2] and Keck[14] chromospheric indices, suggest the presence of active regions on the stellar surface. This can be reconciled with the stability of GJ 436 emission if its spin axis is tilted[8], so that active regions could be frequently occulted by the planet while yielding a small rotational flux modulation. Using 14 years of ground-based differential photometry, we confirm this modulation and derive a stellar rotation period $P_{rot}$ = 44.09±0.08 days, which implies that GJ 436 is older than 4 Gyr (Methods). This value agrees well with the periods of 40.6±2.2 days and 44.5±4.6 days that we derive from periodograms of the Hα and Ca II$_{H\&K}$ activity indicators, respectively. Combining the stellar radius with our results for $P_{rot}$ and $V_{eq} \sin i_\star$ yields $i_\star = 39^{+13}_{-9}°$ (degenerate with $i_\star = 141^{+9}_{-13}°$), confirming the tilt of the star spin axis with respect to the line of sight. By chance these degenerate values for $i_\star$ yield similar distributions for the true 3D obliquity of GJ 436b, which imply a nearly polar orbit with $\Psi_b = 80^{+21}_{-18}°$ (Fig. 2, Methods).

GJ 436b has a puzzling eccentricity, $e_b$ = 0.16[2]: tidal interactions with the star should have circularized its orbit in less than ~1 Gyr[15,4], unless the internal structure of the planet results in abnormally weak tides[4,15,16], or a hypothetical distant companion GJ 436c perturbs its orbit. Circularization could take up to 8 Gyr if GJ 436b and c evolved to a quasi-stationary secular fixed point in which their orbital apses are co-linear[17]. However this scenario requires coplanar orbits in a specific initial configuration, which our measurement of GJ 436b spin-orbit angle disfavors. This misalignment is unlikely to arise from scattering with a companion, as this usually occurs in young systems, and GJ 436b's orbit would have since been circularized. It is also surprising because tides in the thick convective envelope of cool stars are expected to realign close-in planets efficiently[10,18,1]. In fact, there is one other outlier in the low-obliquity systems with short tidal dissipation time-scales[18]: WASP-8b is on an eccentric ($e = 0.3$)[19], misaligned ($\lambda = -143°$)[20] orbit that would take a similar duration to GJ 436b to re-align (Methods). Dynamical interactions with a massive, long-period companion have been proposed[19] to explain the architecture of the WASP-8 system. The eccentricity and obliquity of GJ 436b[4] could originate from a similar Kozai migration induced by a candidate perturber, hereafter called GJ 436c. Fig. 3 shows a migration pathway that could have led to the architecture of the system in ~5 Gyr. In a first phase lasting for ~4 Gyr, GJ 436c induces strong oscillations in the eccentricity of GJ 436b and their mutual inclination, which naturally misaligns the GJ 436b orbital plane. At the onset of the second phase, the orbital distance of GJ 436b and the mutual inclination drop sharply to their present-day value. The mutual inclination keeps oscillating slightly, which results in larger oscillations of GJ 436b 3D obliquity consistent with the measured value. The orbit of GJ 436b, excited to a high eccentricity during the first phase, slowly circularizes and reaches the present value in ~1 Gyr. Different Kozai migrations could have led to the present architecture, and acceptable values for the initial orbit of GJ 436b, the mass and period of GJ 436c can be constrained (Methods) by combining Kozai simulations with RV measurements, direct imaging, and our constraints on the age of the system (4–8 Gyr). We illustrate this approach in Fig. 4, which shows that planetary or brown dwarf companions with masses between ~0.04 and 40 $M_{jup}$ and periods between ~3 and 400 yr could have driven GJ 436b into Kozai cycles if it was initially further than ~ 0.2 au from the star. The subsequent inward migration could have altered the nature of GJ 436b, triggering the atmospheric escape that sustains the giant cloud of hydrogen trailing today the planet[5]. Meanwhile, weak tidal dissipation would have left the orbit of GJ 436c mostly unchanged over time, except for its mutual inclination with GJ 436b. By constraining its present-day value, one could determine the 3D orientation of the GJ 436c orbital plane (Methods, Fig. 2).

Since the reloaded RM technique directly retrieves the intrinsic stellar surface velocity, it can probe the architecture of planetary systems even around cool, slowly-rotating stars. Combining the technique with next-generation infrared spectrographs (SPIRou, NIRPS) will allow for a detailed characterization of the systems discovered around M dwarfs by upcoming transit surveys (CHEOPS, TESS and PLATO), revealing whether GJ 436b is the exception rather than the rule.

**Acknowledgements** Based on observations made with the HARPS spectrograph on the 3.6 m ESO telescope at the ESO La Silla Observatory, Chile, under GTO program ID 072.C-0488, and with the Italian Telescopio Nazionale Galileo (TNG) operated on the island of La Palma by the Fundación Galileo Galilei of the INAF (Istituto Nazionale di Astrofisica) at the Spanish Observatorio del Roque de los Muchachos of the Instituto de Astrofísica de Canarias under OPTICON program 16A/049, "Sensing Planetary Atmospheres with Differential Echelle Spectroscopy" (SPADES). OPTICON has received funding from the European Community's Seventh Framework Programme (FP7/2013-2016) under grant agreement number 312430. This project has also received funding from the European Research Council (ERC) under the European Union's Horizon 2020 Research and Innovation Programme under grant agreement number 724427 (FOUR ACES). This work was carried out in the framework of the National Centre for Competence in Research "PlanetS" supported by the Swiss National Science Foundation (SNSF). R.A., N.A.-D., V.B., D.E, C.L., A.W. acknowledge the financial support of the SNSF. HMC gratefully acknowledges support as a CHEOPS Fellow from the SNSF National Centre of Competence in Research "PlanetS". G.W.H. acknowledges long-term support from Tennessee State University and the State of Tennessee through its Centers of Excellence program. X.B. and X.D acknowledge the support of CNRS/PNP (Programme national de planétologie). X.B. acknowledges funding from the European Research Council under the ERC Grant Agreement n. 337591-ExTrA. We thank C.A. Watson for calculating the convective mass of GJ 436, H. Knutson for facilitating the determination of the stellar rotation period, J-B. Delisle for discussing the system geometry, and the TNG staff for the service observation.

**Author Contributions** V.B. coordinated the study of the GJ 436 system, performed the reduction and analysis of the transit data, interpreted the results, and wrote the paper. V.B. and D.E proposed the idea at the origin of the manuscript. D.E. developed the HARPS-N transit observation programme, which was led by D.E. and A.W. V.B., H.M.C., and C.L. developed and refined the reloaded RM technique. H.B. performed the Kozai simulations and contributed to the interpretation. G.W.H. derived the stellar rotation period from analysis of photometry. N.A.-D. and X.D. derived the stellar rotation period from analysis of activity indices. X.B. and N.A.-D. analysed RVs, and D.S. analysed direct imaging data, used to constrain GJ 436c. R.A., H.C., C.L., and A.W. contributed to the analysis and interpretation of the transit data. All authors discussed the results and commented on the manuscript.

**Author Information** Reprints and permissions information is available at www.nature.com/reprints. The authors declare no competing financial interests. Readers are welcome to comment on the online version of the paper. Correspondence and requests for materials should be addressed to V.B. (vincent.bourrier@unige.ch).

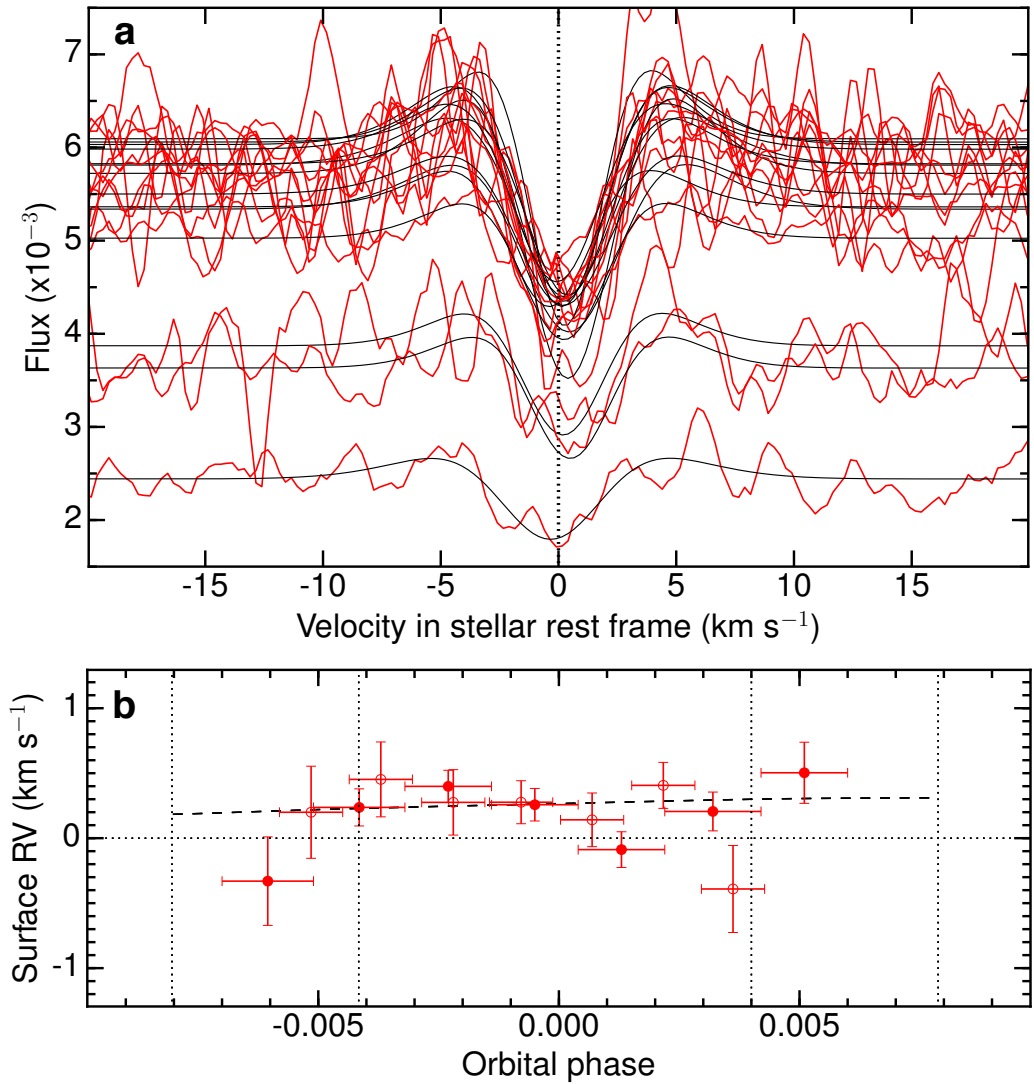

Figure 1 | **Properties of the stellar photosphere along the transit chord of GJ 436b. a,** CCF$_{PO}$ and their double-Gaussian best fits (black lines) as a function of velocity in the star rest frame. Visits 2 and 3, obtained with the same instrument at similar orbital phases, were binned together. Flux level varies with limb darkening and the area occulted by the planet. **b,** Intrinsic radial velocities of the stellar surface (symbols are empty for Visit 1, filled for Visit 2+3) and their best-fit model (dashed line) as a function of GJ 436b orbital phase. Dotted lines are transit contacts. Horizontal bars show the exposure durations. 1$\sigma$ uncertainties are propagated from the continuum dispersion in **a.**

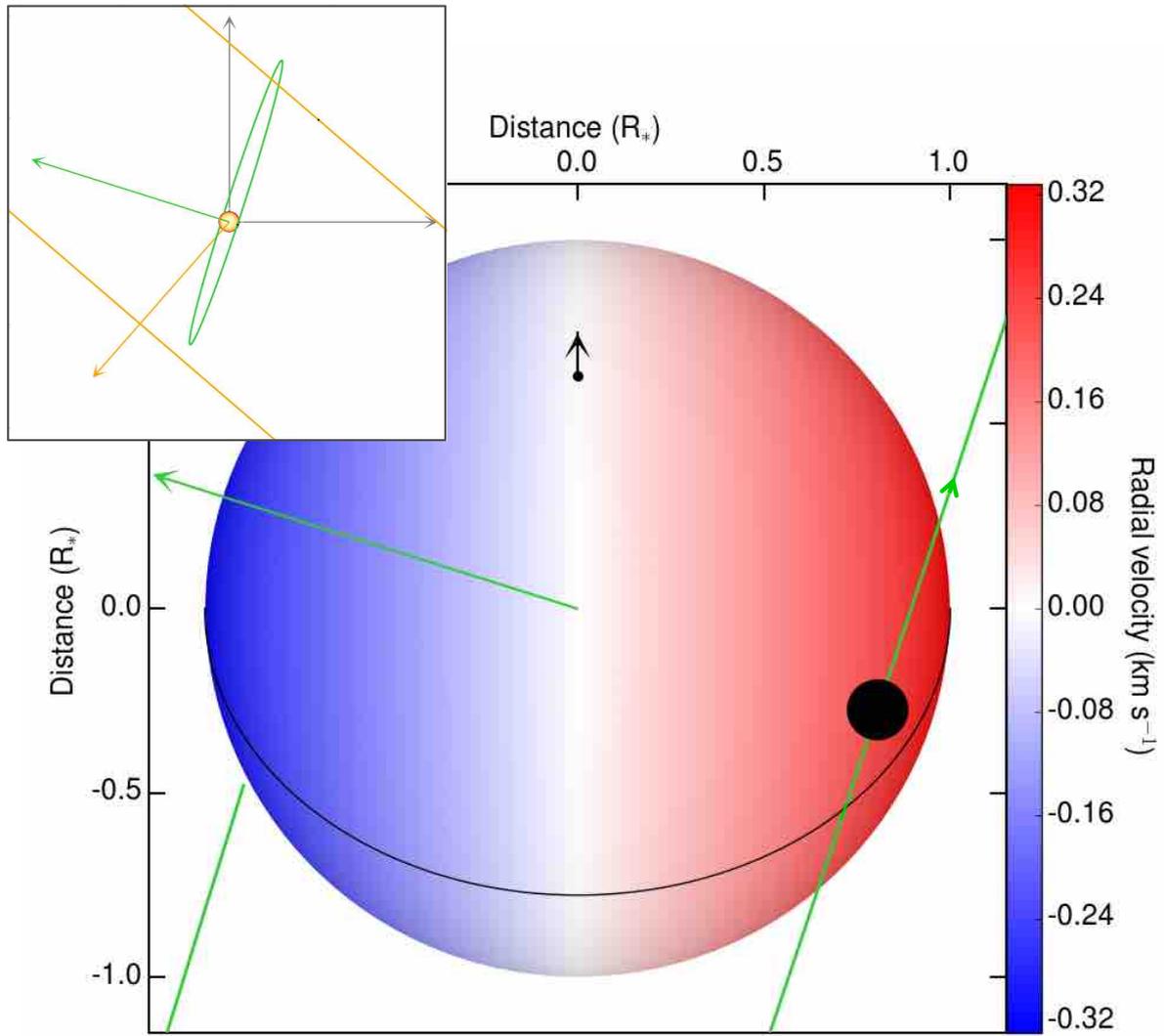

**Figure 2 | Architecture of the GJ 436 system, projected on the plane of the sky.** Stellar disk color corresponds to its surface RV field. The black arrow from south to north pole (visible in this configuration with $i_\star = 39°$) is the inclined stellar spin. A solid black line represents the stellar equator. GJ 436b (black disk) orbital axis is shown as a green arrow of same length as the half stellar spin axis, and its orbital trajectory as a solid green curve. **Inset**, zoom out of this image, showing in orange a possible orbit for the perturber GJ 436c ($i_c = 89.8°$, $\lambda_c = 139°$, $a_c = 7.9$ au; Methods). Grey axes are the sky-projected stellar spin axis and node line.

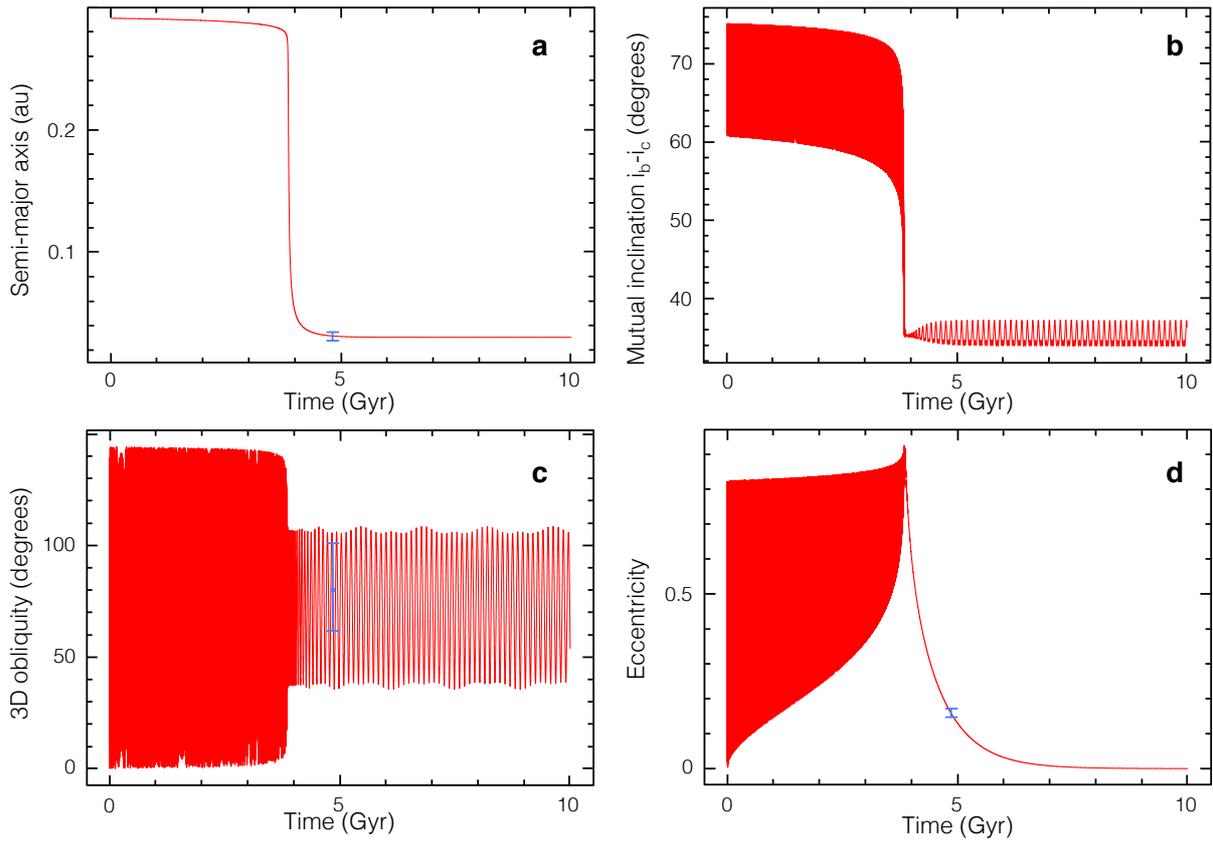

**Figure 3 | Secular evolution of GJ 436b.** Possible Kozai migration pathway that would have led to the present-day architecture of the GJ 436 system in about 5 Gyr, with $M_c$ = 0.23 $M_{jup}$ the mass of GJ 436c and $a_c$ = 7.9 au its orbital distance (Methods). The semi-major axis of GJ 436b **(a)** and its mutual inclination with GJ 436c **(b)** quickly drop once Kozai cycles end, while its eccentricity **(d)** slowly decreases. Low oscillations of the mutual inclination lead to larger variations of the 3D obliquity of GJ 436b orbital plane **(c)**. Blue points correspond to the known properties of GJ 436b.

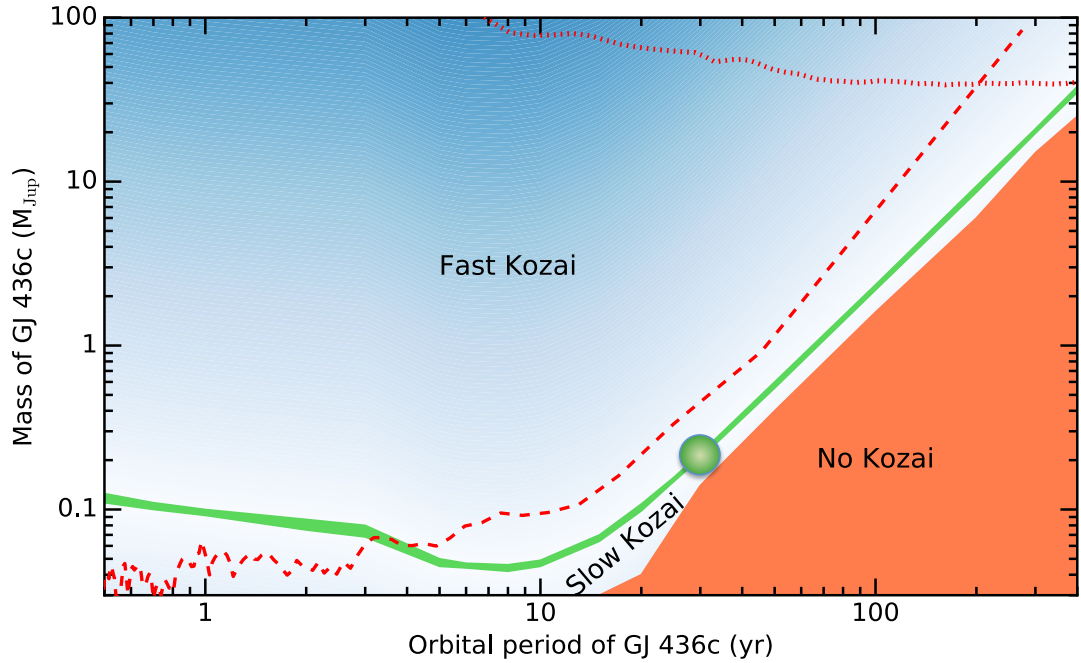

**Figure 4 | Constraints on the mass and period of a putative perturber GJ 436c.** The age of the system constrains the width of the green region, which delimits the properties that would have allowed GJ 436c (green disk) to drive GJ 436b to its present-day orbital configuration via Kozai migration. In the Fast Kozai region, GJ 436b would already be circularized. In the Slow Kozai region, Kozai cycles would still be ongoing. RV measurements and direct imaging exclude regions above the dashed and dotted red lines, respectively (the RV curve is a limit on $M_c \sin i_c$). This diagram shows a subset of possible migrations, for the initial properties of GJ 436b (mutual inclination $i_m^0 = 85°$, $a_b^0 = 0.35$ au) and GJ 436c used in Fig 3.

## METHODS

**Data analysis and correction of systematics**

Our study is based on three transit observations of the exoplanet GJ 436b with ground-based echelle spectrographs. We obtained 77 and 71 exposures of 400 s duration with HARPS-N on 18 Mars and 11 April 2016, respectively, in the frame of the SPADES program (principal investigator: D.E.). These datasets are complemented with 44 archive exposures of 300 s duration, obtained with HARPS on 9 May 2007, which were previously used to attempt a detection of the Rossiter-McLaughlin effect[2]. Observations were reduced with the HARPS (version 3.5) and HARPS-N (version 3.7) Data Reduction Software, yielding spectra with resolution 115,000 covering the region between 380 and 690 nm. The reduced spectra were passed through an order-by-order cross-correlation with a M2-type mask function, weighted by the depth of the lines, to compute the cross-correlation functions (CCFs) defined in the Solar System barycenter rest frame.

The CCFs of GJ 436 display sidelobes typical of M dwarf stars (Extended Data Fig. 1). Single-Gaussian models, or Gaussian plus polynomial models limited to a portion of the CCF RV range[21], do not use the full information contained in such CCFs, which can limit the stability and the precision of their derived properties (RV, full width at half maximum FWHM, and contrast). We pioneer a new model consisting in the sum of a Gaussian function representing the CCF continuum and sidelobes, and an inverted Gaussian function representing the CCF core. This double-Gaussian model fits well the entire CCF profile, yielding low-dispersion residuals between the CCFs and their best fit (Extended Data Fig 1). The RV centroid of the lobe component is redshifted with respect to the core component, but individual exposures show little dispersion around the average redshift in each visit (Extended Data Fig 2). Similarly the ratios between the amplitude of the Gaussian components, and the ratios between their FWHM, are stable in each visit. The properties of the two Gaussian components are thus tightly correlated, and we fixed the RV centroid difference, the amplitude ratio, and the FWHM ratio to their average value in each night, leaving our model with only four free parameters (continuum level; RV centroid, amplitude or contrast, and FWHM of the core Gaussian component).

Earth's atmosphere induces a global variation in the flux measured during a night, leading to the loss of absolute flux levels and variations in the distribution of flux with wavelength that can be different for each exposure. This changes the relative contribution to the CCF of lines that have different width and contrast, but share similar Doppler shifts. Therefore, CCFs uncorrected for the flux unbalance show strong variations in FWHM and contrast over each night, while their RVs are little affected (Extended Data Fig. 3). Visit 1 is more stable than Visits 2 and 3, most likely because GJ 436 culminates close to the zenith when observed with HARPS-N and thus varies strongly in elevation over the night, while it remains at similar low elevations when observed with HARPS. The reloaded RM technique[3] relies on the comparison of the in- and out-of-transit CCFs, and therefore requires a high stability of the CCF profiles over each night. The standard correction of the flux unbalance by the HARPS and HARPS-N pipelines is not applied by default to M dwarfs because their spectra vary considerably with sub-spectral type. We thus applied a correction customized to GJ 436. For each exposure in a given night, we integrate the flux between 1/4 and 3/4 of each order in the 2D extracted spectra (ie, the flux at the top of the blaze function). This yields low-resolution spectra defined as a function of the central wavelength in each order. We create a template by combining several low-resolution spectra selected for their high signal-to-noise ratio. All low-resolution spectra are normalized, divided by the template, and fitted with a sixth-order polynomial. The original 2D spectra for each exposure are divided by the corresponding best-fit polynomial, before recalculating their CCFs. The corrected $CCF_{DI}$ show a very stable contrast and FWHM (Extended Data Fig. 3), and their RVs show little dispersion around the Keplerian curve calculated with GJ 436b known orbital properties (Extended Data Table 2). In each night, some exposures have signal-to-noise ratios too low in their bluest orders to be corrected, and are excluded from our analysis (Extended Data Table 1).

**Application of the reloaded RM technique**

$CCF_{DI}$ are corrected for the Keplerian motion of the star[2] and shifted into the star rest frame using the systemic velocities derived from double-Gaussian fits to the $CCF_{DI}^{OT}$ (9.79 km s$^{-1}$ in all visits). Compared to Doppler tomography[22], which pioneered the direct analysis of planet-induced distortions in the stellar CCFs but assumes constant photospheric line profiles, the reloaded RM technique enables a cleaner isolation of the planet velocity-space trajectory across the star through the analysis of the local $CCF_{PO}$ obtained by subtracting the in-transit $CCF_{DI}$ from the $CCF_{DI}^{OT}$. Since the absolute flux level of the $CCF_{DI}$ is lost, they were calibrated photometrically using GJ 436b transit light curve calculated with the *batman* package[23]. We used non-linear limb-darkening coefficients derived from transit photometry of GJ 436b in a visible band[24] covering most of HARPS and HARPS-N spectral range (Extended Data Table 2). $CCF_{PO}$ are assigned flux errors set to the standard deviation in the flat region of their continuum. Since CCFs are oversampled by the instrument pipelines (steps of 0.25 km s$^{-1}$ for a pixel width of 0.82 km s$^{-1}$), we measured the standard deviation after removing three in four points. Uncertainties on the parameters derived from the double-Gaussian fits to the CCF$_{PO}$ are 1σ statistical errors from a Levenberg-Marquardt least-squares minimization. We assumed that all CCFs of GJ 436 share similar double-Gaussian profiles, i.e. that the RV between the lobe and core components of the CCF$_{PO}$, the ratio between their amplitude, and the ratio between their width, were set to the average values derived from the fits to the CCF$_{DI}$ in each visit. This assumption was validated *a posteriori* by the good fit of this model to the CCF$_{PO}$, with no spurious features found in the residuals.

While the shape of the transit light curve must be known to apply the reloaded RM technique, the orbital properties and ephemeris of GJ436b could potentially be derived from the analysis of the surface RVs. However, these properties are determined more precisely through photometry and velocimetry than through analysis of the RM effect[9], and were thus fixed to the values in Extended Data Table 2. Nonetheless, we varied each of these parameters within their 1σ uncertainties and confirmed that the associated surface RVs never deviated beyond the 1σ uncertainties of the nominal values in Fig. 1.

**Analysis of the stellar surface velocity field**

Under the assumption of solid-body rotation (reasonable for mid-M dwarfs[25]), $V_{eq}$ and $i_\star$ are degenerate because the analysis of the surface RVs alone does not allow the determination of the stellar latitudes transited by the planet. We thus fitted $\lambda_b$ and $V_{eq} \sin i_\star$ with the reloaded RM model[3] using uniform priors in a custom-made MCMC algorithm[9]. We applied an adaptive principal component analysis so that step jumps take place in an uncorrelated space, which better samples the posterior distributions. We analysed the system with multiple chains, started at random points in the parameter space. We checked that all chains converged to the same solution, thinned them using the maximum correlation length of the parameters, and merged them to obtain posterior distributions with a sufficient number of independent samples. The best-fit values for the model parameters are set to the medians of the posterior probability distributions and their 1σ uncertainties are evaluated by taking limits at 34.15% on either side of the median (Extended Data Fig. 6).

GJ 436 passed close to the zenith in Visits 2 and 3, which can lead to tracking issues with the HARPS-N telescope (TNG) due to its altazimuth mount. This occurred much earlier than the transit in Visit 2 (near phase -0.049), with no apparent negative effects on our results (Extended Data Fig. 3). In Visit 3, TNG staff astronomers reported tracking issues with exposures at phases 0.0031 and 0.0052. GJ 436 culminated just after phase 0.0031 (elevation 87.85°), and exposures on both sides were also taken close to the zenith with elevations of 87.49° (phase 0.0014) and 87.17° (phase 0.0052). Thus, fiber injection issues might have affected the three last in-transit exposures in Visit 3 (Extended Data Fig. 5), which could explain the two RV deviations observed at phases 0.0014 and 0.0031. However, the RV of the last exposure at phase 0.0052 is consistent with the best-fit model and with the other visits, and the contrast and FWHM of these three last in-transit exposures show no deviations compared to the other visits. Finally, the largest of the three RV deviations in Visit 3 comes from the first $CCF_{PO}$ during ingress, which is faint and might thus yield less accurate measurement. Since the origin of these RV deviations is not clear, and they do not substantially influence the derived best-fit model, we kept them in our analysis.

**Rotation period and age of GJ 436**

We observed GJ 436 during 14 seasons between 2003 November and 2017 May with the T12 0.80m Automatic Photoelectric Telescope (APT) at Fairborn Observatory in Arizona[26]. This yielded 1986 measurements in Strömgren *b* and *y* photometric pass bands, combined into a single pass band to improve the precision (~1.5–2.0 mmag for a single observation). We computed differential magnitudes of GJ 436 versus the mean brightness of two comparison stars (HD102555 and HD103676), which were constant to 1 mmag during all observing seasons. Extended Data Fig. 7 shows the nightly differential magnitudes after observations in the transit window were removed. Observations were corrected for long-term variations and normalized so each observing season has the same mean, yielding an overall dispersion of 4.1 mmag. We performed a frequency analysis based on least-squares sine fits with trial periods between 1 and 100 days. The goodness of fit at each period is measured as the reduction factor in the variance of the data, yielding a clear detection at 44.09±0.08 days. The uncertainty is derived from the FWHM of the peak associated with this photometric period, which we interpret as the stellar rotation period $P_{rot}$ made evident by rotational modulation in the visibility of surface starspots (Extended Data Fig. 7). Five out of the 14 individual seasons show definitive periodic variations in agreement with $P_{rot}$, ranging from 41.7 to 46.6 days with a weighted mean of 44.44±0.30 days.

We used our measurement of $P_{rot}$ to constrain the age of GJ 436, estimated at 3.7±3.9 Gyr by its observed effective temperature and bolometric flux[27]. Observations of cool stars in open clusters show that stellar rotation periods increase with redder color (lower mass). Stars in the 2.5 Gyr old cluster NGC 6819[28] have a lower spin-down rate for $B-V > 0.65$, the period increasing from 19 to 23 days when *B-V* increases from 0.65 to 0.88. Since this rate is not expected to increase with lower masses, we can extrapolate that GJ 436 (*B-V* = 1.45) would rotate in 33 days maximum if it was 2.5 Gyr old, showing that it is in fact much older. Cool stars in the open cluster M67, aged 4.2 Gyr, show a similar flattening of the spin-down rate[29] for *B-V* > 0.65, the period increasing from about 25–30 to 30–35 days when *B-V* increases from 0.65 to 1. With $P_{rot}$ = 44 days the age of GJ 436 is likely close to 4–5 Gyr, and we consider 4–8 Gyr to be a conservative range.

**Inclination of the star spin axis and 3D obliquity of GJ 436b**

We combined our measurement of the period and the stellar projected rotational velocity to derive the inclination of the star spin axis $i_\star = \arcsin[P_{rot} \, V_{eq} \sin i_\star /(2\pi \, R_\star)]$, with $R_\star$ the stellar radius . It is then possible to determine the 3D obliquity between the normal to GJ 436b orbital plane and the spin axis of the star, $\Psi_b = \arccos(\sin i_\star \cos \lambda_b \sin i_b + \cos i_\star \cos i_b)$, with $i_b$ the orbital inclination of GJ 436b. To determine best-fit values and uncertainties for $i_\star$ and $\Psi_b$ we sampled randomly their probability distributions, assuming a Gaussian distribution for $P_{rot}$ and using the MCMC probability distributions obtained for $V_{eq} \sin i_\star$ and $\lambda_b$. There remains a degeneracy between the star spin axis pointing toward or away from the observer, yielding $i_\star = 39^{+13}_{-9}°$ or $i_\star = 141^{+9}_{-13}°$. Nonetheless, because of the high projected obliquity the corresponding values for $\Psi_b$ ($77^{+20}_{-15}°$ or $82^{+19}_{-15}°$) are compatible with each other. We consider their average, $80^{+21}_{-18}°$, as the 3D obliquity of the system.

**Tidal dissipation timescale of GJ 436b**

We placed the GJ 436 system in Fig. 4 of ref. 18, which shows obliquity measurements as a function of $\tau = (M_b/M_{conv})^{(-1/3)}(a/R_\star)$, a quantity proportional to the mass of the stellar convective envelope ($M_{conv}$) and to the scaled distance to the star ($a_b/R_\star$), and thus to the tidal dissipation timescale (where $M_b$ is the mass of GJ 436b). We derived $M_{conv} \sim 0.146$ for GJ 436, using the `EZ_WEB` stellar evolution code (http://www.astro.wisc.edu/~townsend/static.php?ref=ez-web ; this results is largely insensitive to the age of the star and its initial mass). Fig. 4 in ref. 18 shows that systems with short tidal dissipation time-scales ($\tau \leq 700$) are preferentially aligned ($\lambda \leq 20°$). The two only outliers in this distribution of low-obliquity systems are GJ 436 ($\tau \sim 180$, $\lambda_b = 72°$) and WASP-8 ($\tau \sim 240$, $\lambda = 143°$).

**Kozai migration of GJ 436b**

The Kozai migration of GJ 436b was presented and simulated with a *N*-body + tides code in ref. 4. We show in Fig. 3 a possible evolution based on our new constraints on the system. The semi-major axis of GJ 436b had to be initially larger than today, to prevent tidal effects circularizing its orbit too fast. During a first phase GJ 436c induced strong oscillations of the eccentricity and inclination of GJ 436b. At peak eccentricity, inclination and periastron are minimal, and tidal friction slowly decreases the semi-major axis. The bottom eccentricity of the Kozai cycles gradually increases, until it reaches the peak eccentricity and the cycles stop. The orbital distance of GJ 436b and its mutual inclination with GJ 436c then drop sharply because of tides, while the eccentricity of GJ 436b (excited to high values at the onset of the second phase) decreases slowly to its present value. Kozai cycles in the first phase misaligned the orbit of GJ 436b (initially assumed to be within the stellar equatorial plane), leading to strong oscillations of its 3D obliquity. During the second phase the orbit of GJ 436b remains misaligned, and its 3D obliquity keeps oscillating at a slower rate between about 40–105°, in agreement with the measured $\Psi_b$.

Kozai migration primarily depends on the mass $M_c$ and semi-major axis $a_c$ of the perturber GJ 436c, the initial semi-major axis $a_b^0$ of GJ 436b, and the parameter $h^0=|\cos i_r^0|\sqrt{1-(e_b^0)^2}$ (with $e_b^0$ the initial eccentricity of GJ 436b, and $i_r^0$ its mutual inclination with GJ 436c). Our goal is not to explore the full parameter space, but to show that Kozai migration can explain the architecture of the system with no need for an abnormal tidal dissipation factor for GJ 436b, which was thus set to a Neptune-like value of $10^5$ (ref. 4). We used the age of the system (4–8 Gyr) to constrain the transition time $t_{tr}$ between the two phases of the Kozai evolution. This transition time delimits three regions in the ($a_c$, $M_c$) plane (Fig. 4): the "fast Kozai" region ($t_{tr} < 4$ Gyr), excluded because GJ 436b would be circularized today; the "slow Kozai" region ($t_{tr} > 8$ Gyr), excluded because the Kozai cycles would still be ongoing today; the "convenient" region, which allows GJ 436b to be in the later stages of the second phase within the age range of the system. For a given set of initial properties ($a_b^0$, $h^0$), the convenient region thus defines the acceptable values of ($a_c$, $M_c$) for GJ 436c, upon which we can further place upper limits derived from RV measurements and adaptive optics imaging (see next section). We find that the present system architecture can be explained if GJ 436b initially satisfied $a_b^0 \gtrsim 0.2$ au and $h^0 \lesssim 0.17$ (i.e., $i_r^0 \gtrsim 80°$ for small $e_b^0$). In that case, the Kozai migration could have been driven by perturbers with masses between ~0.04 and 40 $M_{jup}$ and periods between ~3 and 400 yr (Fig. 4). We note that other migration pathways exist, different initial conditions for GJ 436b shifting the width and position of the convenient region in the ($a_c$, $M_c$) plane. Future RVs and direct imaging measurements will refine the constraints on these properties.

**Conditions on GJ 436c orbital trajectory**

The mutual inclination between the orbital planes of GJ 436b and GJ 436c satisfies $\cos i_r = \cos i_b \cos i_c + \cos \Omega \sin i_b \sin i_c$, with $i_b$ and $i_c$ the inclinations of the orbital planes, and $\Omega = \omega_c - \omega_b$ the difference between the longitudes of their ascending nodes. Since $i_c$ is known to a high precision, the values satisfying this relation follow the 3D surface shown in Extended Data Fig. 8. If the mutual inclination $i_r$ is known, this relation reduces to an oval ring in the ($\Omega, i_c$) plane. Furthermore, if we take the sky-projected node line of the star as a reference for the longitude of the ascending node $\omega$, the sky-projected obliquity of an orbiting body satisfies $\lambda = \omega$ or $\lambda = \omega - 180°$. It is then possible to constrain the alignment of GJ 436c with $\lambda_c = \lambda_b + \Omega$ or $\lambda_c = \lambda_b + \Omega - 180°$. Constraints on the mutual inclination would thus allow a full determination of GJ 436c orbital trajectory. This will require a complete exploration of Kozai migration pathways, beyond the scope of this study. Here, we illustrate this point with the scenario shown in Fig. 3, where the mutual inclination oscillates between 66 and 68° and constrains $|i_c - 90°| \leq 71°$, $|\Omega| \leq 68°$, and $\lambda_c$ in [-20°, 173°] or [-200°, -6°]. A possible trajectory for GJ 436c is shown in Fig. 2, where we selected $i_r = 67°$ and $i_c = 89.8°$, yielding $\Omega = 67°$ and $\lambda_c = 139°$. The semi-major axis $a_c = 7.9$ au was derived from Fig. 4.

We note that two transiting Earth-sized companions have been postulated in the GJ 436 system[30], on shorter and larger orbits than GJ 436b. However they were not confirmed by later analyses[2], and could not have driven the Kozai migration of GJ 436b given the results of our simulations[4] (Extended Data Fig 5), and the constraints on their properties derived from RV measurements[2] and transit studies[31,32,33].

**Constraints on GJ 436c from RV measurements and direct imaging**

We derived conservative detection limits on $M_c \sin i_c$ from the residuals of HARPS[2] and Keck[14] RV time-series using the same approach as in ref 2. Perturbers above the red line in Fig. 4 are excluded for a given period with a 99% confidence level. We note that the constraint on the true mass of GJ 436c depends on its orbital inclination, which could be derived as explained in the previous section.

We retrieved on the ESO archive (program 081.C-0430; PI: D. Apaï) publicly available high contrast imaging data of GJ 436 taken at the VLT with the Nasmyth Adaptive Optics System (NAOS) Near-Infrared Imager and Spectrograph (CONICA) instrument. The data was taken in April 2008 in the $L'$ band, using the field tracking mode of NACO, no coronagraph, and no image saturation. We used the Geneva High Contrast Imaging Data Reduction Pipeline[34] to reduce the data and compute the $\delta L'$ band detection limits. Since no $L'$ photometry could be found in the literature for GJ 436, we estimated it using near-IR photometry and stellar evolutionary models. We used the low-mass star models of ref. 35 at an age of 5 Gyr and with a solar metallicity, 2MASS $J$, $H$ & $Ks$ apparent magnitudes and the Hipparcos parallax. We obtained a mid-IR magnitude estimate $L'$=5.78±0.03 for GJ 436, which corresponds to a mass of $M_\star$= 0.46 $M_\odot$ and an effective temperature of $T_{\text{eff}}$= 3,610 K, in good agreement with the spectroscopic analysis[27] (Extended Data Table 2). The absolute $L'$-band detection limits as a function of the projected separation is obtained by combining the results of the NACO images and the magnitude estimate of GJ 436 while the conversion into companion's mass detection limits is done using ref. 36 evolutionary models for cool brown dwarfs. Fig. 4 shows that the presence of massive brown dwarfs ($M > 40$ $M_{\text{Jup}}$) at long periods ($P > 90$ yr) is ruled out.

**Code availability**. We have opted not to make available the codes used for the data extraction and analysis as they are currently an important asset of the researchers' tool kits.

**Data availability**. All spectra used in this study are publicly available on the ESO archive (HARPS; (http://archive.eso.org/eso_archive_main.html) and on the Telescopio Nazionale Galileo archive (HARPS-N; http://archives.ia2.inaf.it/tng/). Source Data for Fig. 1 are available online. The other data sets generated and analysed during the present study are available from VB (vincent.bourrier@unige.ch) on reasonable request.

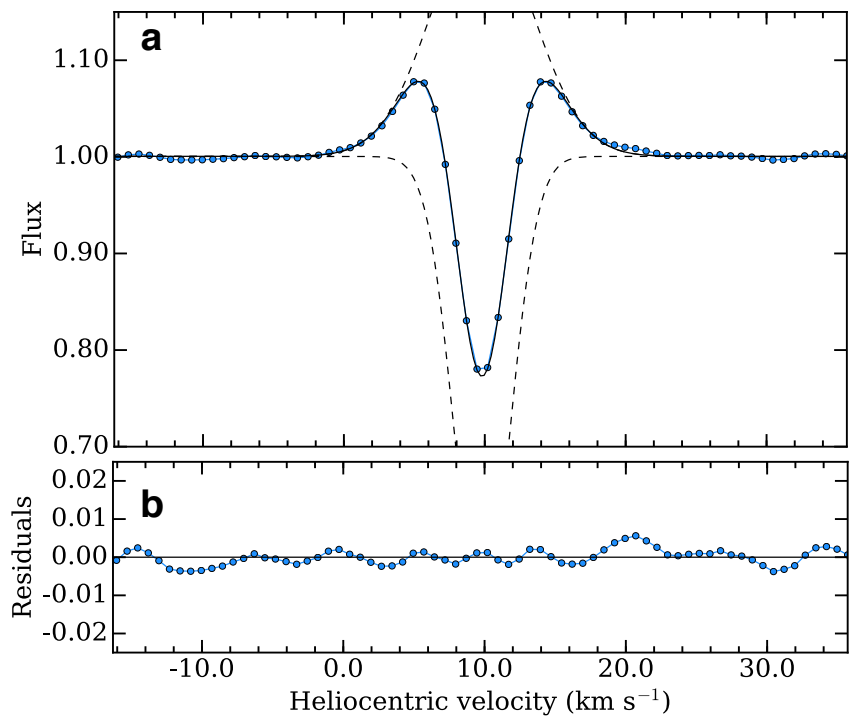

**Extended Data Figure 1 | Observed and modelled CCF of GJ 436. a,** Typical HARPS-N CCF of GJ 436 (blue points), fitted with a double-Gaussian model (solid black line). This model is the combination of a Gaussian profile for the CCF continuum and lobes plus an inverted Gaussian profile for the CCF core (individual components are plotted as dashed black lines). **b,** Residuals between the observed CCF and its best fit.

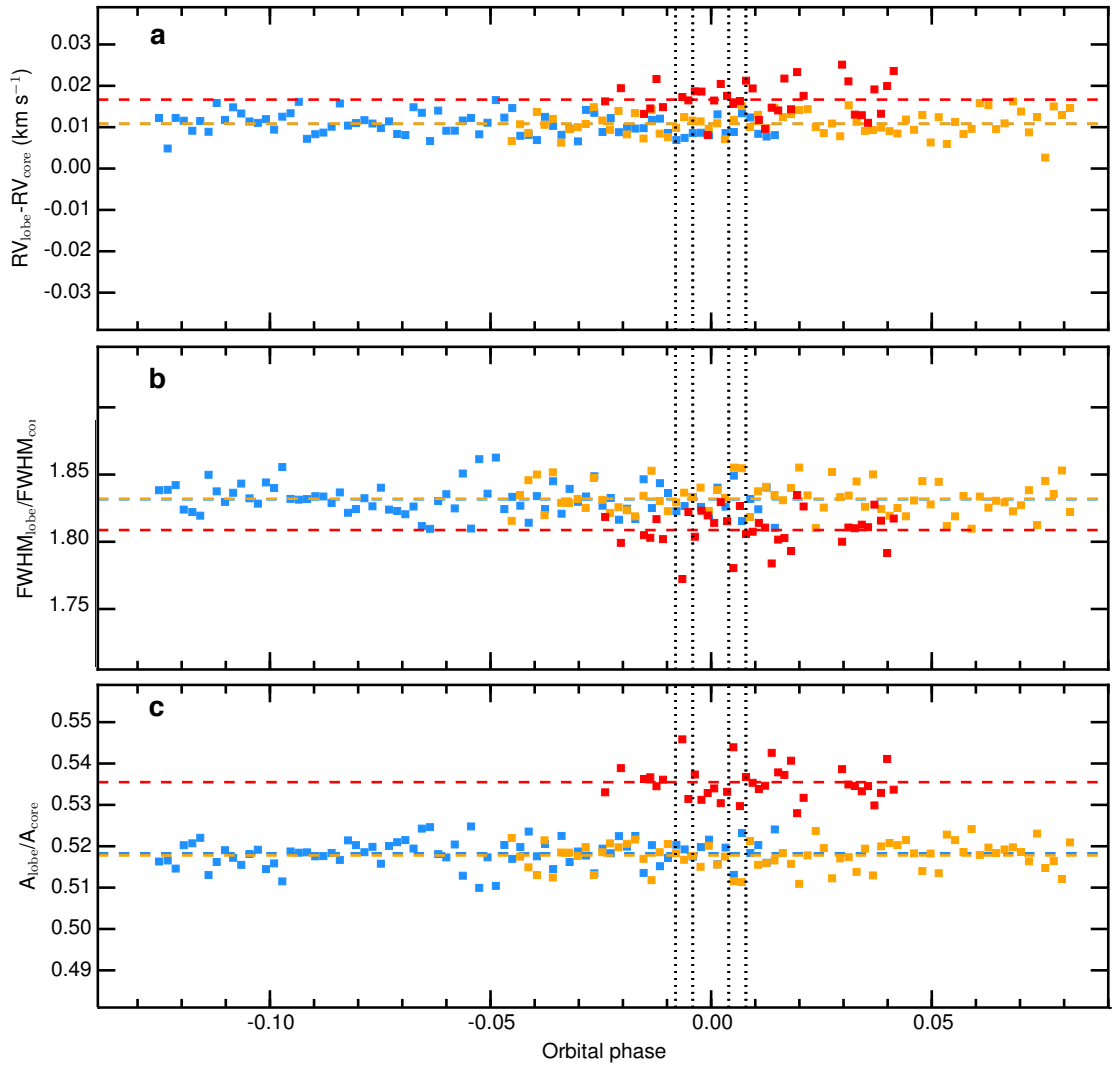

**Extended Data Figure 2 | Comparison between the properties of the lobe and core Gaussian components of the CCF model.** The panels show the difference between the RV centroids of the lobe and core components (**a**), the ratio between their FWHM (**b**), and the ratio between their amplitude (**c**), as a function of GJ 436b orbital phase for each exposure in Visit 1 (red), Visit 2 (blue) and Visit 3 (orange). There is little dispersion of these values around their average in each Visit, shown as dashed horizontal lines. Vertical dotted lines are the transit contacts.

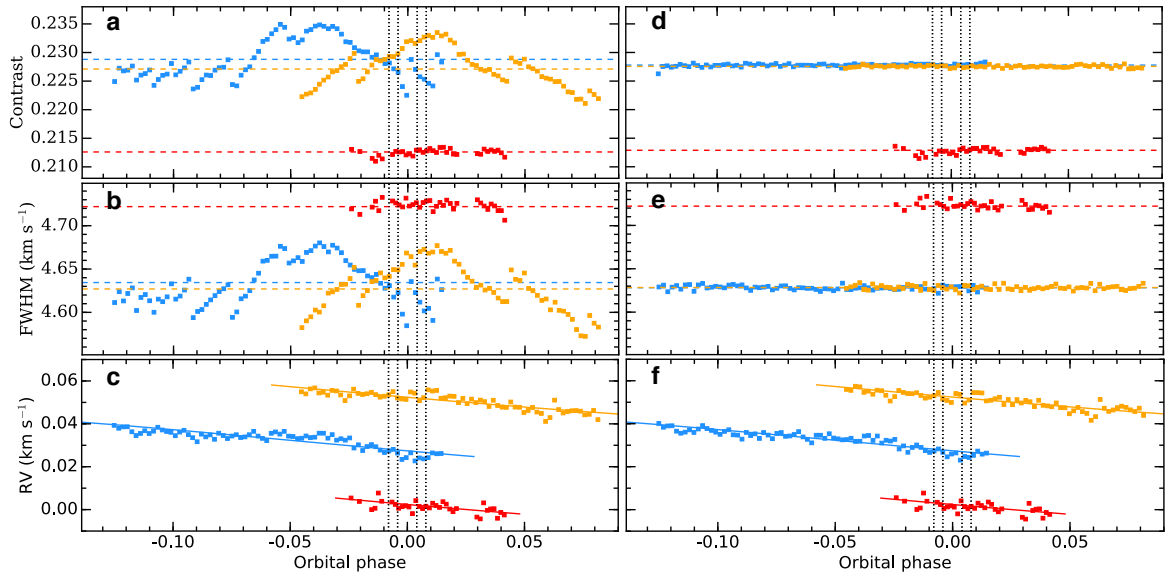

**Extended Data Figure 3 | Correction for the effects of Earth atmosphere.** Properties derived from the double-Gaussian fits to the $CCF_{DI}$ are shown before correction (**a-c**) and after correction of the flux distribution (**d-f**), as a function of GJ 436b orbital phase. The contrast of the $CCF_{DI}$ is shown in **a** and **d**, their FWHM in **b** and **e**, their RV in **c** and **f**. RVs are relative to the systemic velocity in each visit, and have been offset by 25 m s$^{-1}$. They are overplotted with the expected Keplerian RV curve. Visits 1, 2, 3 are colored in red, blue, orange, respectively. Vertical dotted lines are the transit contacts ; horizontal dashed lines show the average values in each visit.

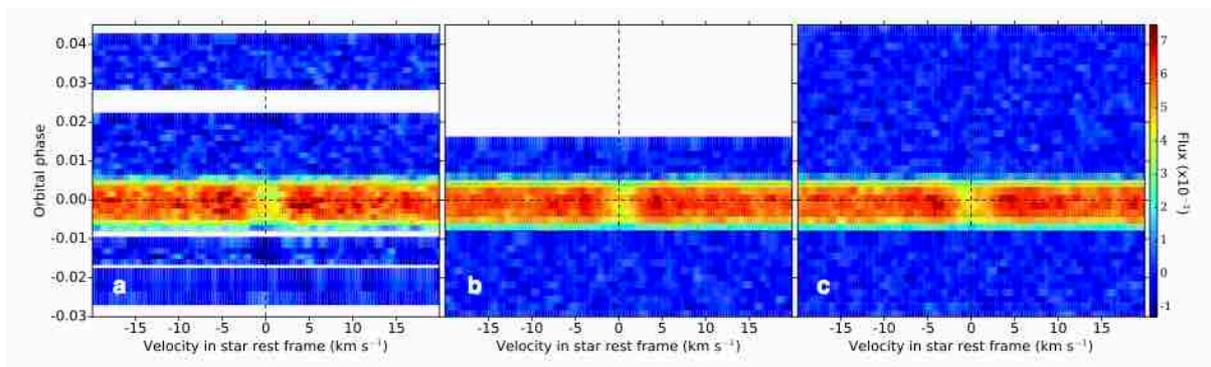

**Extended Data Figure 4 | Maps of the residuals between the scaled $CCF_{DI}$ and the $CCF_{DI}^{OT}$.** Residuals are colored as a function of their flux, and plotted as a function of radial velocity in the stellar rest frame (in abscissa) and orbital phase (in ordinate) for Visit 1 (**a**), Visit 2 (**b**), and Visit 3 (**c**). The vertical and horizontal dashed black lines indicate the mid-transit time and stellar rest velocity, respectively. In-transit residuals correspond to the $CCF_{PO}$, and show the average stellar line profile (recognisable by a lower flux in the $CCF_{PO}$ cores) from the regions occulted by GJ 436b across the stellar disk. Out-of-transit residuals show little dispersion in all visits, consistently with the low activity of the host star.

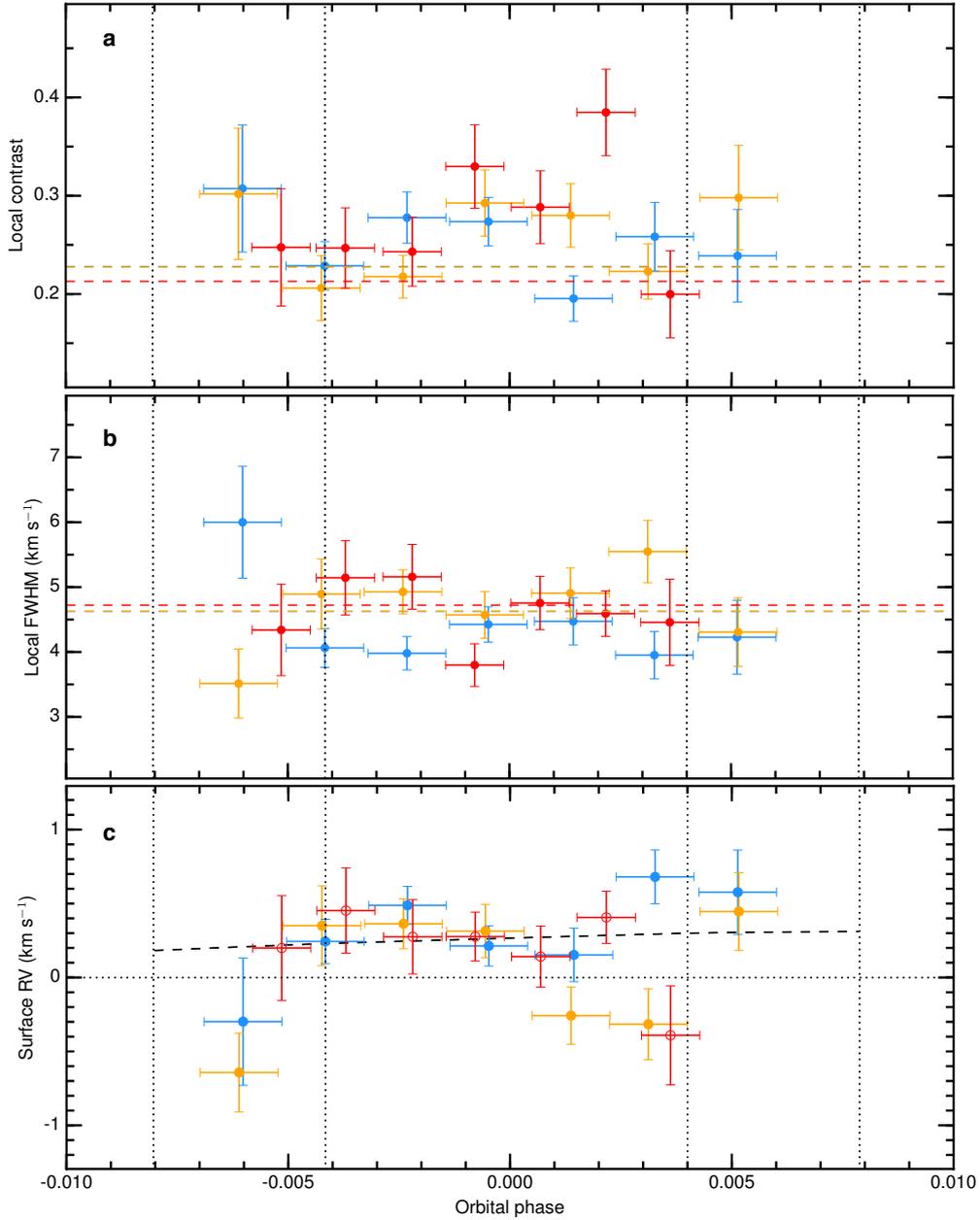

**Extended Data Figure 5 | Properties of the CCF$_{PO}$ as a function of GJ 436b orbital phase.** The contrast (**a**), FWHM (**b**), and RVs (**c**) are derived from the double-Gaussian best fits to the CCF$_{PO}$, and show similar values over the three nights. **a–c,** Visits 1, 2, and 3 are colored in red, blue, and orange respectively. All error bars are 1$\sigma$. Horizontal error bars correspond to the exposure time. Vertical dashed lines are the transit contacts. **a, b,** The width and contrast of the CCF$_{DI}^{OT}$ (horizontal dashed lines) are similar over the three visits. **c,** The dashed black line is the reloaded RM model corresponding to the best-fit for the planet trajectory and the velocity field of the star.

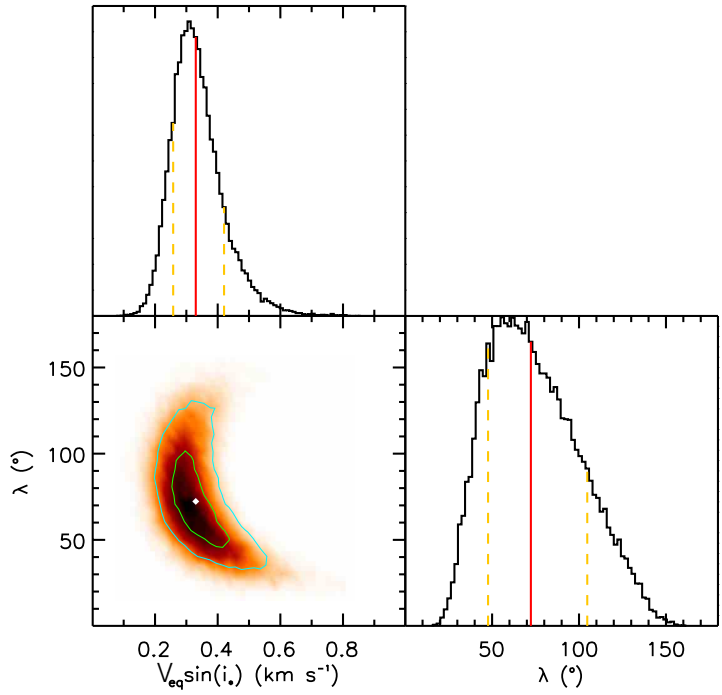

**Extended Data Figure 6 | Correlation diagram for the posterior probability distributions of the solid-body rotation model parameters.** Green and blue lines show the 2D confidence regions that contain 39.3% and 86.5% of the accepted steps, respectively. One-dimensional histograms correspond to the distribution projected on the space of each line parameter, with the orange dashed line limiting the 68.3% confidence interval. The red line and white point show median values.

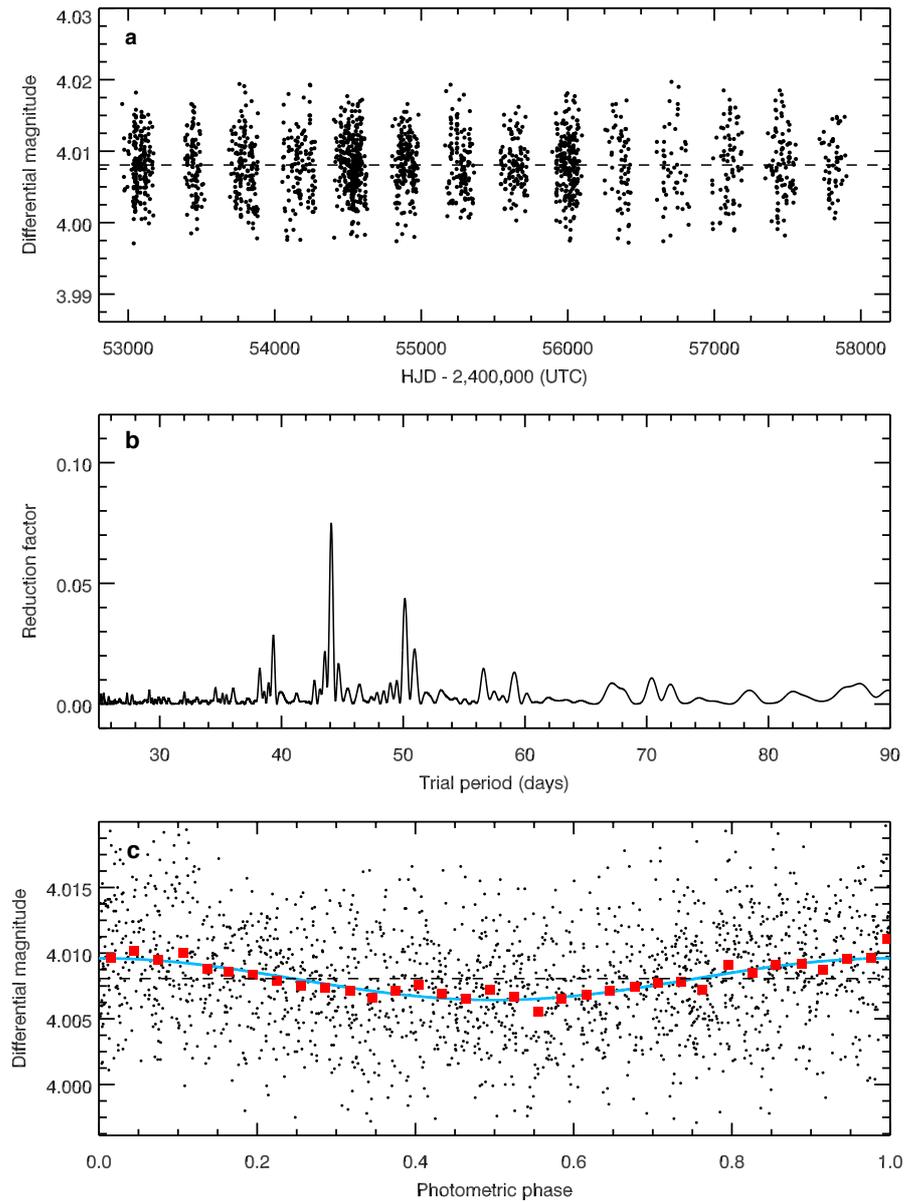

**Extended Data Figure 7 | Ground-based photometry of GJ 436. a,** Time series of GJ 436 nightly magnitude with transit points removed and normalized to the same seasonal mean. UTC, Coordinated Universal time; HJD, heliocentric Julian date. **b,** Frequency spectrum of the normalized observations with strongest peak at a photometric period of 44.09 days, and secondary peaks corresponding to yearly aliases caused by the temporal sampling. **c**, Normalized data and best-fit sine curve (blue line) phased to $P_{rot}$ = 44.09 days. The binned data (red squares) highlights the low-level brightness modulation of GJ 436 (peak-to-peak amplitude of 0.0032 mag)

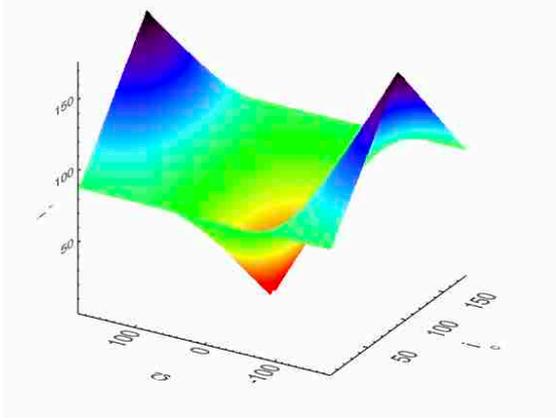

**Extended Data Figure 8 | Conditions on GJ 436b and GJ 436c orbital planes.** For a given mutual inclination $i_r$ (vertical axis), the acceptable properties for the orbital planes describe an oval ring in the ($\Omega$, $i_c$) plane. $\Omega$ is the difference between the longitudes of the ascending nodes, and $i_c$ the orbital inclination of GJ 436c.

**Extended Data Table 1 | Log of GJ 436b transit observations**

| Visit number | 1 | 2 | 3 |
| --- | --- | --- | --- |
| Observation date | 9 May 2007 | 18 March 2016 | 11 April 2016 |
| Instrument | HARPS | HARPS-N | HARPS-N |
| Number of exposures | 44 | 77 | 71 |
| Exposures kept after color-correction | 35 | 76 | 69 |
| Before transit | 6 | 63 | 20 |
| During transit | 11 | 9 | 9 |
| After transit | 18 | 4 | 40 |
| Orbital phase of exposures that failed our color-correction | -0.018; -0.017; -0.009; -0.008; 0.022; 0.024; 0.025; 0.027; 0.028 | 0.016 | 0.083; 0.085 |
| Orbital phase of exposures that failed our CCF detection criterion | -0.0065; 0.0050; 0.0065; 0.0079 | -0.0079; 0.0070 | -0.0079; 0.0069 |

**Extended Data Table 2 | Properties of the GJ 436 system**

| Name | Fixed properties | Value | Reference |
| --- | --- | --- | --- |
| Stellar radius | $R_*$ | 0.449±0.019 $R_\odot$ | Ref. 27 |
| Stellar mass | $M_*$ | 0.445±0.044 $M_\odot$ | Ref. 27 |
| Effective temperature | $T_{eff}$ | 3479±60 K | Ref. 27 |
| Non-linear limb-darkening coefficients | $u_1$ | 1.47 | Ref. 24 |
| | $u_2$ | -1.10 | Ref. 24 |
| | $u_3$ | 1.09 | Ref. 24 |
| | $u_4$ | -0.42 | Ref. 24 |
| Semi-major axis | $a_b/R_*$ | 14.54±0.14 | Ref. 2 |
| Mid-transit time | $T_b$ | 2454865.084034±0.000035 BJD | Ref. 2 |
| Orbital period | $P_b$ | 2.64389803±2.6x10$^{-7}$ days | Ref. 2 |
| Orbital inclination | $i_b$ | 86.858+0.049-0.052° | Ref. 2 |
| RV semi amplitude | $K_b$ | 17.59±0.25 m s$^{-1}$ | Ref. 2 |
| Planet mass | $M_b$ | 25.4+2.1-2.0 $M_{Earth}$ | Ref. 2 |
| Transit depth | $(R_b/R_*)^2$ | 0.006819±0.000028 | Ref. 2 |
| Eccentricity | $e_b$ | 0.1616±0.004 | Ref. 2 |
| Argument of periastron | $\omega_b$ | 327.2+1.8-2.2° | Ref. 2 |
| Name | Derived properties | Value | |
| Projected rotational velocity | $V_{eq} \sin i_*$ | 0.330+0.091-0.066 km s$^{-1}$ | |
| Projected obliquity | $\lambda_b$ | 72+33-24° | |
| Stellar rotation period | $P_{rot}$ | 44.09±0.08 days | |
| Stellar inclination | $i_*$ | $39^{+13}_{-9}$° | |
| | | $141^{+9}_{-13}$° | |
| 3D obliquity | $\psi_b$ | $80^{+21}_{-18}$° | |